\documentclass[usenatbib]{mn2e}
\usepackage{graphicx}

\title{Stellar populations in the surrounding field of the LMC clusters
NGC~2154 and NGC~1898}
\author[Chiosi et al.]
{E. Chiosi$^{1}$,G. Baume$^{2}$, G. Carraro$^{3,5}$, E. Costa$^{4}$, A. Vallenari$^{1}$, R. A. M\'endez$^{4}$\\
$^{1}$ INAF, Osservatorio Astronomico di Padova, Vicolo Osservatorio 2, I-35122, Padova, Italy\\
$^{2}$ Facultad de Ciencias Astron\'omicas y Geof\'isicas (UNLP), Instituto de Astronom\'ia de la Plata (CONICET,UNLP),\\  Paseo del Bosque s/n, La Plata, Argentina\\
$^{3}$ ESO, Alonso de Cordova 3107, Santiago de Chile, Chile\\
$^{4}$ Departamento de Astronom\'ia, Universidad de Chile, Casilla 36-D, Santiago, Chile\\
$^{5}$ Dipartimento di Astronomia, Universit\'a di Padova, Vicolo Osservatorio 3, I-35122, Padova, Italy}
\begin{document}
\date{Received .../ Accepted ...}

\maketitle
\begin{abstract}
In this paper we present a study and comparison of the star formation rates 
(SFR) in the fields around NGC~1898 and NGC~2154, two intermediate-age
star clusters located in very different regions of the Large
Magellanic Cloud.
We also present a photometric study of NGC 1898,
and of seven minor clusters which happen to fall in the field of
NGC 1898, for which basic parameters were so far unknown.
We do not focus on NGC~2154, because this cluster 
was already investigated in Baume et al. 2007, using the same theoretical tools.
The ages of the clusters were derived by means of the isochrone fitting method on their 
$clean$ color-magnitude diagrams. Two distinct populations of clusters were found: one cluster 
(NGC~2154) has a mean age of 1.7 Gyr, with indication of extended
star formation over roughly a 1 Gyr period,
while all the others have ages between 100 and 200 Myr. The SFRs 
of the adjacent fields were inferred using the downhill-simplex algorithm. Both SFRs show
enhancements at 200, 400, 800 Myr, and at 1, 6, and 8 Gyr. 
These bursts in the SFR
are probably the result of dynamical interactions between the Magellanic Clouds (MCs), and of the
MCs with the Milky Way.
\end{abstract}


\section{INTRODUCTION}

Stellar clusters have traditionally been considered as the natural cradle of stars, from which 
stars can migrate into the field \citep{lada2003}. We cannot exclude however the possibility 
that stars also form in the field itself \citep{bressert2010}. To evaluate the statistical 
relevance of stars formed in-situ among field populations, it is necessary to make detailed 
comparative studies (including ages) of many clusters and their related fields. The Magellanic 
Clouds (MCs) are located close to the Galaxy (about 60 Kpc),  their members can be considered 
essentially equidistant, and they present a rich population of clusters. Therefore they 
provide an ideal laboratory to address this matter.\\

In this paper we present an analysis of the star formation rate (SFR) in the $fields$ around
two populous stellar clusters of the LMC: NGC~1898, located in the central LMC bar, and
NGC~2154, located in its NE border as shown in Fig.1. We also present a photometric study of NGC 1898,
and of seven minor clusters which happen to fall in the field of NGC 1898, for 
which basic parameters were so far unknown. We note that NGC~2154 was subject of a previous
study by our group, using the same theoretical tools \citep[hereafter Bau07]{baume2007}.\\

The above sample is particularly interesting for it comprehends both disk and bar populations that are
recognized to have different star formations histories (SFHs), as discussed by Vallenari et al 1994
and Harris et al 2009.  \citet{holtzman1999} find, based on HST data, that there
is a significant component of stars older than 4 Gyr in the outer fields and in the bar. They
also notice that there is no age gap in the field SFR, unlike the case of the cluster SFR. Olsen
et al. (1999) analyze six fields in the LMC and find that all of them have significant recent 
($<3-4 Gyr$) star formation. They find that bar fields experience more star formation in the range
of 4-8 Gyr than disk fields. Among other works we mention a recent study on the LMC star formation history
 by \citet{rubele2012} based on VISTA data. They analyze the SFH in several regions of the LMC. In particular
we identify tile 6-6 as the region of NGC~1898 and tile 8-8 as the region of NGC~2154. They find a continuous SF in
the bar (NGC~1898) while some burst can be recognized in the disk area (NGC~2154) at 20 Myr, 1-2 Gyr and 8 Gyr.\\  

\begin{figure*}
\centering
\resizebox{\hsize}{!}{\includegraphics{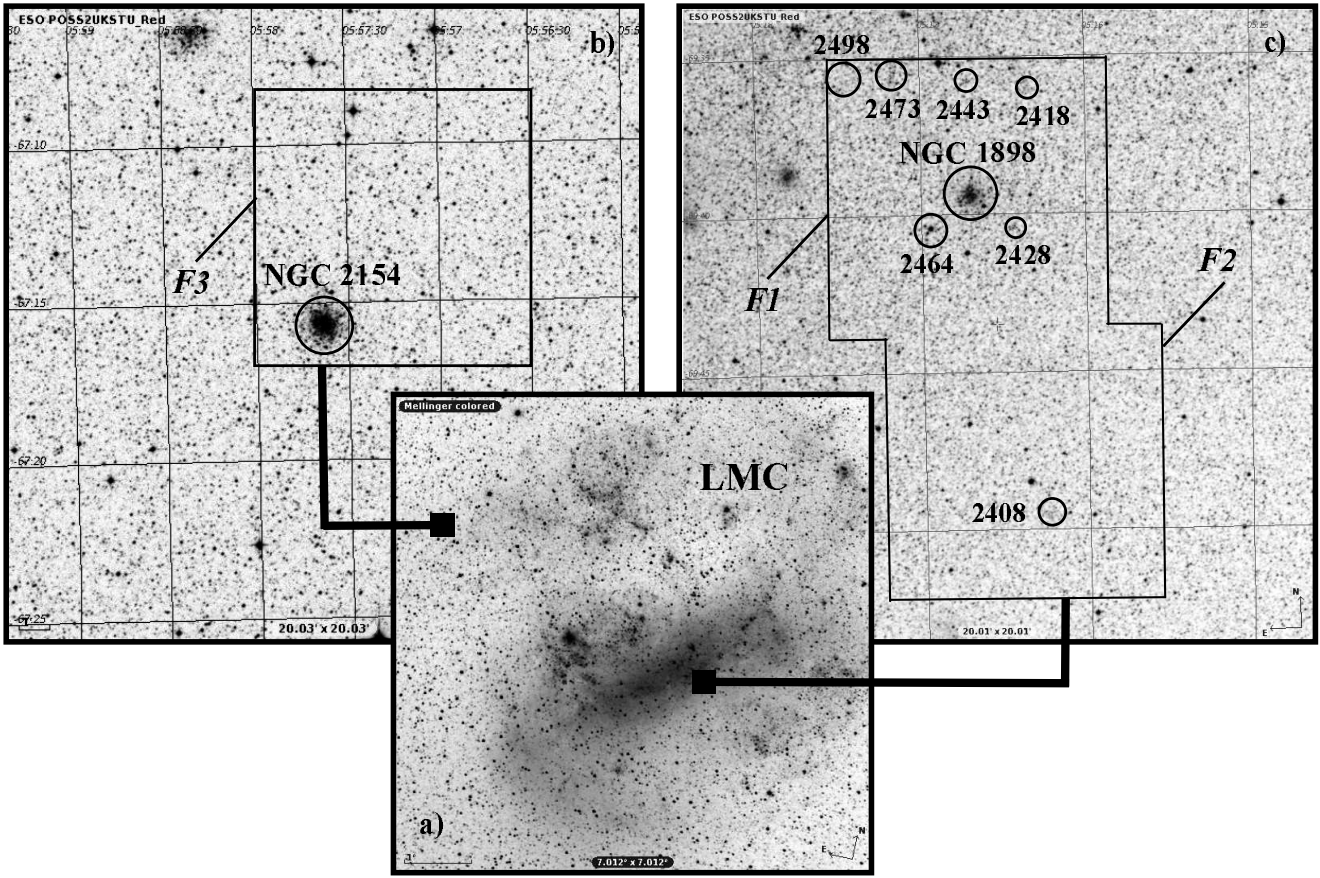}}
\caption{Upper panels b) and c) show the $8\farcm85 \times 8\farcm85$ areas surveyed by our $BR$ 
observations ($F1$, $F2$ and $F3$), superimposed on approximately $20\farcm0 \times 20\farcm0$
DSS-2 red images. We note that fields $F1$ and $F2$ have a very small overlap ($\sim$15$arcsec$),
which is not noticeable in panel c) . Circles in them depict the location, and
approximate size of the star clusters studied. North is at the top and East is to the left.
Lower panel (a) shows the location of our fields in the LMC.}
\label{fig:dss}
\end{figure*}
          
The paper is structured as follows: in section 2 we summarize the observations and the data 
reduction procedures, in section 3 we describe the methods used for the star formation rate,
 in section 4 the main results for the two fields and in section 5 we present the cluster analysis.
 Finally, in section 6, we
summarize the conclusions of our analysis.

\section{OBSERVATIONS AND DATA REDUCTION}
In the followings we describe the observation and data reduction procedure for the NGC~1898 field. 
For the equivalent description of the field of NGC~2154 we refer to Bau07.\\

Our study is based on $B(R)_{KC}$ observations carried out using a 24$\mu$ pixel Tektronix 2048 
$\times$ 2048 detector attached to the Cassegrain focus of the du Pont 2.5-m telescope at Las 
Campanas Observatory, Chile. Gain and read noise were 3 e-/ADU and 7 e-, respectively. This set-up 
provided direct imaging over a field of view (FOV) of $8\farcm85 \times 8\farcm85$ with a scale of 
0.259 arcsec/pix. This relatively large FOV allowed us to study a good sample of the LMC field 
population around the clusters. A log of the observations is given in Table~1. Seeing 
varied between 0.9 and 1.5 arcsec. A log of the observations for NGC~2154 can be found in Table~1 of 
Bau07.\\

The observations presented is this paper were secured as a part of a large survey to study the
SFH and absolute proper motion of the MCs \citep{noel2007,noel2009,costa2009,costa2011}.\\

All frames were processed using standard IRAF tasks. We used the CCDRED package for the pre-reduction
procedure, for which purpose zero exposures and sky flats were taken every night. PSF instrumental
magnitudes were obtained in the standard way using the DAOPHOT package (Stetson 1987), and the
DAOMASTER code (Stetson 1992) was used to combine the corresponding photometric tables for different
exposure times and/or filters. Our instrumental photometry was calibrated using the PHOTCAL package,
for which purpose we observed several $UBVRI$ standard star areas (Landolt 1992; namely fields Mark A,
PG0231+051, PG2213-006, SA098, SA110, SA113 and TPhe), and performed aperture photometry on them.\\

Our transformation equations were:\\

\begin{equation}
b = B + b_1 + b_2 X + b_3 (B-R)
\end{equation}

\begin{equation}
r = R + r_1 + r_2 X + r_3 (B-R)
\end{equation}

In these equations $b$ and $r$ are the instrumental magnitudes normalized to 1 sec, and $X$ is 
the airmass. The values of the transformation coefficients in the above equations are also listed
in Table~1. The rms of the fits in the blue and red bands turned out to be 0.026 and 0.036,
respectively.\\

\begin{table}
\begin{center}
\begin{tabular}{ccccc}
\hline
$Field$ & $Filter$ & $Date$ & $Airmass$ & $Exptime$      \\
        &          &        &           & ($times N \ sec$) \\
\hline
    &     & 2007-10-10 & 1.37 & 60       \\
F1  & $B$ & 2007-10-10 & 1.37 & 800      \\
    &     & 2008-10-28 & 1.36 & 6 x 800  \\
\hline
    &     & 2007-10-10 & 1.39 & 120      \\
F1  & $R$ & 2007-10-10 & 1.38 & 600      \\
    &     & 2007-10-06 & 1.36 & 10 x 450 \\
\hline
    &     & 2007-10-10 & 1.34 & 60       \\
F2  & $B$ & 2007-10-10 & 1.34 & 800      \\
    &     & 2008-10-29 & 1.35 & 8 x 800  \\
\hline
   &      & 2007-10-10 & 1.36 & 120      \\
F2 &  $R$ & 2007-10-10 & 1.35 & 600      \\
   &      & 2007-10-09 & 1.35 & 17 x 500 \\
\hline
\hline
\multicolumn{5}{l}{Transformation Coefficients} \\
\hline
\multicolumn{3}{l}{$b_1 = 1.058 \pm 0.025$}  & \multicolumn{2}{l}{$r_1 = 0.613 \pm 0.049$} \\
\multicolumn{3}{l}{$b_2 = 0.213 \pm 0.019$}  & \multicolumn{2}{l}{$r_2 = 0.136 \pm 0.038$} \\
\multicolumn{3}{l}{$b_3 = -0.043 \pm 0.005$} & \multicolumn{2}{l}{$r_3 = -0.009 \pm 0.007$} \\
\hline
\end{tabular}
\end{center}
\caption{Logfile of observations in the area of NGC 1898, together with the coefficients used in
our transformation equations (1) and (2).  $N$ indicates the number of images obtained for each
integration time.}
\label{tab:log}
\end{table}

\begin{figure}
\centering
\resizebox{\hsize}{!}{\includegraphics{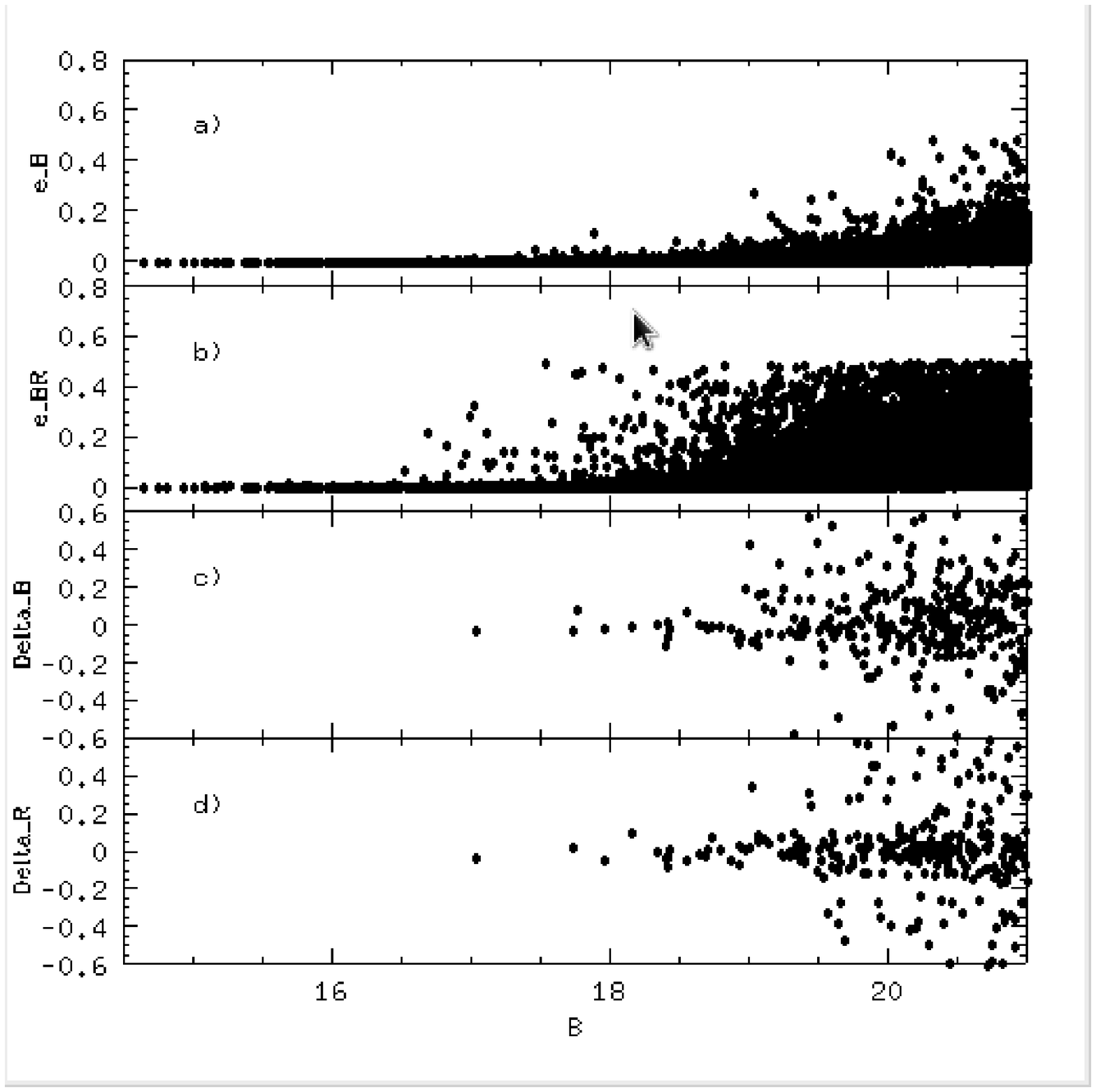}}
\caption{Panels a) and b): photometric errors (from DAOPHOT) in $B$ and $(B-R)$, plotted as
a function of $B$ magnitude. Panels c) and d): comparison of the photometry in fields $F1$ and
$F2$, based on their overlapping region}
\label{fig:errors}
\end{figure}

Our photometric errors (from DAOPHOT) in $B$ and $(B-R)$ are plotted as a function of $B$
magnitude in panels a) and b) of Fig.~\ref{fig:errors}. Given that there is a very small overlap
between fields $F1$ and $F2$ (not obvious in Figure~1), it was possible to compare the photometry
secured in each of them.  The differences $Delta_B$ and $Delta_R$ as a function of $B$ are
plotted in panels c) and d) of Fig.~\ref{fig:errors}.

\subsection{COMPLETENESS}
The definition of photometric completeness is of fundamental importance for the
clusters analysis and for the reconstruction of stellar populations in the determination of star formation history (SFH).
 The procedure for completeness determination is well described in Bau07 and substancially is based on the injection in the
 image of a suitable number of artificial stars of known magnitude and position. The number of recovered artificial stars
 through the complete data reduction pipeline over the initial number per magnitude interval consitutes the completeness
 coefficient for that interval. We summarize the main results for completeness in the NGC~2154 and NGC~1898 cluster and field
 areas in Table \ref{tab:comp}.

\begin{table}
\begin{center}
\begin{tabular}{l r r r r }
\hline
Delta B   &    Cluster &  Field & Cluster & Field\\
          &    NGC2154 &  NGC2154&NGC1898 & NGC1898\\
\hline
16.0-16.5 &            &         &  100\% &  100\%\\ 
16.5-17.0 &            &         &  100\% &  100\%\\ 
17.0-17.5 &            &         &   97\% &  100\%\\ 
17.5-18.0 &            &         &   93\% &  100\%\\ 
18.0-18.5 &            &         &   83\% &  100\%\\ 
18.5-19.0 &            &         &   74\% &  100\%\\ 
19.0-19.5 &            &         &   59\% &  100\%\\ 
19.5-20.0 &     100\% &  100\%   &     55\% &  100\%\\ 
20.0-20.5 &      93\% &  100\%   &     56\% &  100\%\\ 
20.5-21.0 &      75\% &  100 \% &     42\% &   96\%\\  
21.0-21.5 &      57\% &   100\% &     30\% &   86\%\\  
21.5-22.0 &      57\% &   100\% &     31\% &   77\%\\  
22.0-22.5 &      56\% &   100\% &     23\% &   62\%\\  
22.5-23.0 &      55\% &   100\% &     24\% &   53\%\\  
23.0-23.5 &      54\% &   83\%   &     16\% &   40\%\\ 
23.5-24.0 &      42\% &   61\%   &     12\% &   33\%\\ 
24.0-24.5 &      29\% &   62\%    &      8\% &   20\%\\ 
24.5-25.0 &      32\% &   67\%  &              & \\
25.0-25.5 &      35\% &   82\%  &              & \\
25.5-26.0 &      36\% &   50\%&                & \\
\hline
\end{tabular}
\end{center}
\caption{Completeness study for the NGC2154 region and NGC1898 region; cluster and field.}
\label{tab:comp}
\end{table}

\section{FIELD STAR FORMATION RATE}\label{fieldSFR}
\subsection{METHODS}\label{methods}

In order to derive the field SFR in the regions of NGC~2154 and NGC~1898, we must first subtract the 
clusters from the corresponding fields.
In the case of NGC~2154, we removed the area within a radius of 500 pixels from the cluster center. 
This area extends well outside the core radius of NGC~2154 ($a = 14\farcs7$, about 57 pixels), as 
was determined from the fit with Elson profiles made by Bau07. In the case of NGC~1898 a radius of 
300 pixels was adopted.\\
     
The field SFR is then obtained comparing the observed field CMD with synthetic CMDs, by means of a 
minimization algorithm.\\

The first step consists in creating  a set of synthetic stellar populations, and a grid to be applied 
to the field CMDs. The former were created using the Bertelli ZVAR code release,
 based on the \citet{girardi2000} 
set of evolutionary tracks.\\

The above code needs a set of parameters which must be tuned to the specific case in question. In 
particular, we need to assume an age-metallicity relation and an initial mass function (IMF). The 
age-metallicity relation was adopted from \citet{pagel1998} and is summarized in Table~\ref{tab:age_met}. 
The adopted IMF was that of \citet{kroupa2002}, which is a power law function with a slope of $x = 2.3$, 
for stellar masses $M > 0.5 M_{\odot}$, and of $x = 1.3$ for the 0.08 - 0.5 $M_{\odot}$ mass range.\\

\begin{table}
\begin{center}
\begin{tabular}{c  l }
\hline
$Age Interval$ & $Metallicity$ \\
$[years]$      & $Z$           \\
\hline
6.3e7 : 2e8    & 0.010 \\
2e8   : 3e8    & 0.010 \\
3e8   : 4e8    & 0.010 \\
4e8   : 5e8    & 0.010 \\
5e8   : 6e8    & 0.007 \\
6e8   : 8e8    & 0.007 \\
8e8   : 1e9    & 0.007 \\
1e9   : 2e9    & 0.005 \\
2e9   : 3e9    & 0.004 \\
3e9   : 4e9    & 0.004 \\
4e9   : 5e9    & 0.003 \\
5e9   : 6e9    & 0.003 \\
6e9   : 8e9    & 0.003 \\
8e9   : 1e10   & 0.002 \\
1e10  : 1.2e10 & 0.002 \\
\hline
\end{tabular}
\end{center}
\caption{Age-metallicity relation adopted in the ZVAR code for our LMC
fields. From Pagel \& Tautvaisiene (1998).}
\label{tab:age_met}
\end{table}

We generate populations of 12000 stars for each age interval, covering a range of ages from a few 
Myr to 10 Gyr. The stars were distributed according to the IMF from brightest to faintest, down to the 
magnitude limit set by completeness.\\
     
The grids applied to the CMDs (see Figs. 4 and 5: GRID 2154 and GRID 1898) were built in a way that 
enhances the most important evolutionary stages. A fine binning was used along the main sequence in 
order to resolve the different turn-offs of the contributing populations, while a coarser division 
was adopted for the red clump and sub-giant branch regions. This last feature reflects the uncertainties 
resulting from both the experimental procedure and the theoretical models.\\

A key point in the simulation is the completeness and photometric error reproduction. Both aspects were 
considered in the algorithm to generate the synthetic populations, as explained in the next paragraph. 
They result to be particularly critical in the case of NGC~1898 where the photometric errors are larger 
at brighter magnitudes than in the case of NGC~2154. We notice that the 50\% completeness
limit settles at $B\simeq23$ corresponding to the turn-off of a population of 
6.3 Gyr. The determination of the SFH for ages older than this limit is therefore largely uncertain
and degenerate
in the case of NGC~1898 but older populations are needed to fill the red clump bin.
 In the case of NGC~2154, we refer the reader to Bau07 for a discussion of the
photometric errors and completeness. We note that for this cluster the completeness limit 
 settles
at $B\simeq25.5$ well below the turn-off of a population of 10Gyr.\\


Having created the synthetic populations and the grid, we are in condition to generate histograms 
(i.e. number of stars falling in each sector of the CMD), for both the single theoretical stellar 
populations and for the data. At this stage, we need to introduce photometric errors in the stellar 
models and apply a completeness correction to the theoretical populations. We also need to apply a
reddening correction, and adopt a distance modulus for the LMC. Following Westerlund (1997), we have 
used a reddening of $E_{B-V}=0.08$ and a distance modulus of $B_0-M_B=18.5$.\\

Finally, we determine the best set of coefficients weighting the theoretical histograms that best 
reproduce the observational histogram. To this aim we made use of the downhill simplex method of 
optimization (Nelder \& Mead 1965). The downhill simplex acts as a probe moving in a $N+1$ parameter 
space, where $N$ is the number of theoretical populations and therefore of coefficients. Its shape in 
the $N+1$ parameter space is defined by $N+1$ initial points. It starts calculating at a given point 
the $\chi^2$ function set by the sum of the squared differences between the corresponding bins of the 
theoretical histogram resulting from the given mixture and the observational histogram. Then it moves
to another point through reflection and again calculates $\chi^2$, and so on, resizing and reflecting 
it can define a gradient of the $\chi^2$ and following this gradient it rapidly converges to a minimum. 
To prevent settling on local rather than global minima, 30000 random directions are searched for a new 
minimum. More details can be found in \citet{chiosi2007}.\\

\begin{figure}
\centering
\resizebox{\hsize}{!}{\includegraphics{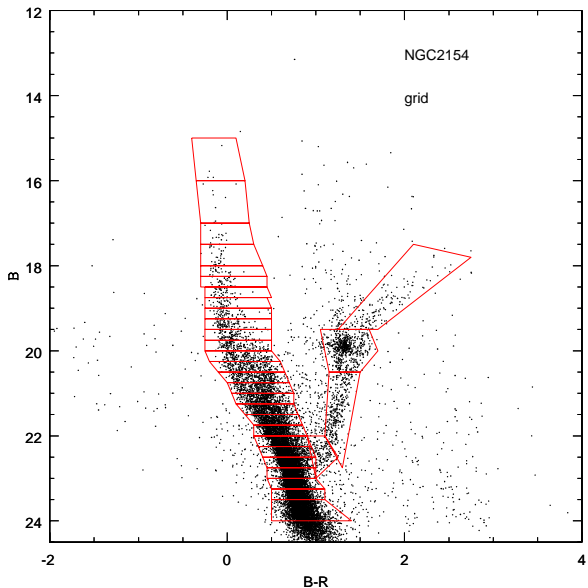}}
\caption{CMD of the region observed in the vicinity of NGC~2154, with the superimposed grid}
\label{fig:ngc2154grid}
\end{figure}

\begin{figure}
\centering
\resizebox{\hsize}{!}{\includegraphics{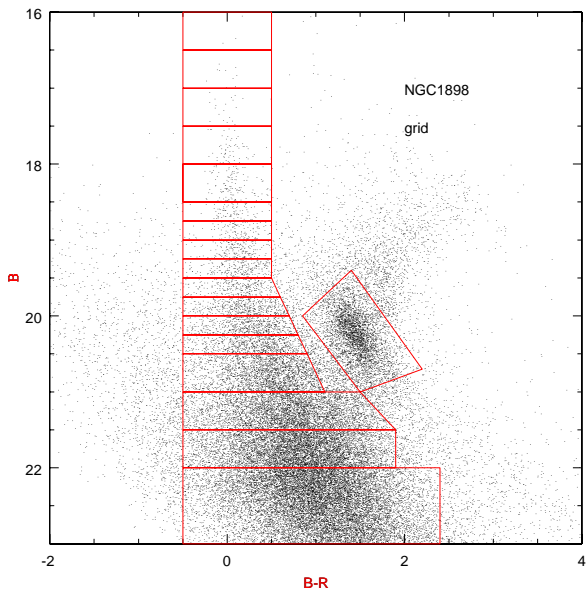}}
\caption{CMD of the region observed in the vicinity of NGC~1898, with the superimposed grid}
\label{fig:ngc1898grid}
\end{figure}

\section{THE HISTORY OF STAR FORMATION IN THE FIELD} \label{THE HISTORY OF STAR FORMATION IN THE FIELD}

\subsection{NGC~2154}

NGC~2154 is located in the NE border of the LMC ($\alpha = 5^h 57^m 38^s$; $\delta = -67^{\circ} 15' 42''$; 
see Figs. 1 and 12 of \citet{costa2009}). There are no previous determination of the SFH in this region.\\  

As shown by Fig.~\ref{fig:ngc2154sfr}, our results indicate that the SFH in the field of NGC 2154 had 
bursts of star formation at 100-200 Myr, 400 Myr, 1-2 Gyr, 6 Gyr and 10 Gyr. We note that Olsen (1999), 
who studied the field SFR in six regions located mainly in the LMC bar, found bursts of star formation 
at 1 Gyr, 5 Gyr, and at ages older than 10 Gyr. These bursts of star formation are probably the result 
of dynamical interactions between the LMC and the SMC, at 200 Myr, and of the MCs with the Milky Way
(MW), at 1.5 Gyr \citep[see]{bekki2005,murai1986}. 
We note however that the validity of this statement cannot be tested
at present because, given the uncertainties of the available proper motions measurements for the MCs
(see e.g. Costa et al. 2009, 2011), their space motions are not precisely known (and hence
the epochs of their peri-galactic passages and their past binding status -- see \citep{piatek2008}).\\

Our results also show that, in agreement with those of Olsen (1999), the well known gap in the cluster
formation rate in the LMC, between 3-10 Gyr \citep[see]{balbinot2010,geisler1997}, does not apply to
the SFR of its field.\\

\begin{figure}
\centering
\resizebox{\hsize}{!}{\includegraphics{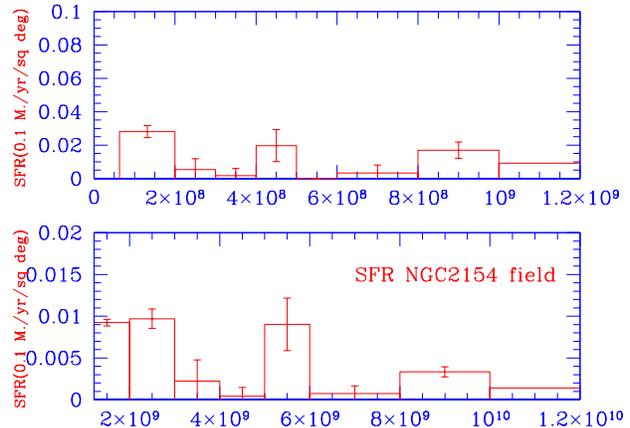}}
\caption{SFR in the field around NGC~2154.}
\label{fig:ngc2154sfr}
\end{figure}

The age of NGC~2154 (1.7 Gyr, from Bau07), falls inside one of the peaks of the field SFH (1-2 Gyr). 
This is in agreement with Subramaniam (2004), who found that star and cluster formation rates in the
LMC are anti-correlated in the age range 30-100 Myr; and correlated in the age range 300-1000 Myr,
and for ages of more than 1 Gyr.\\

As a check of the downhill-simplex algorithm we run the star formation program on the cluster area and 
try to recover the age determination by Bau07. The result is that we find a main SF epysode from 1 to 2 Gyr
coincident with the previous determination plus a peak at 400 Myr probably due to the fact that the field population
is not  subtracted.\\
\begin{figure}
\centering
\resizebox{\hsize}{!}{\includegraphics{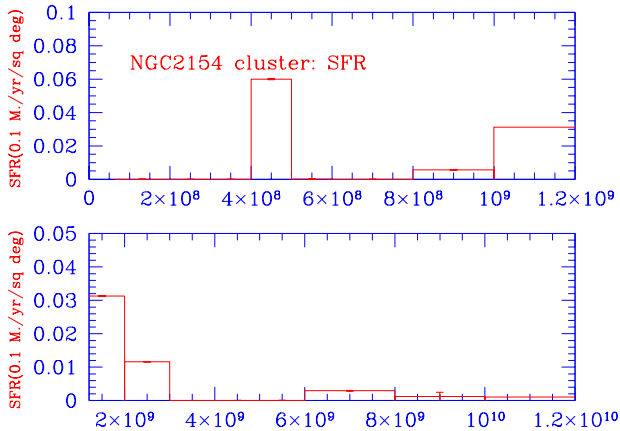}}
\caption{SFR for the cluster NGC~2154.}
\label{fig:ngc2154sfr}
\end{figure}

\begin{figure*}
\centering
\hspace{0.2cm}
\parbox{7.5cm}{\resizebox{7.5cm}{!}{\includegraphics{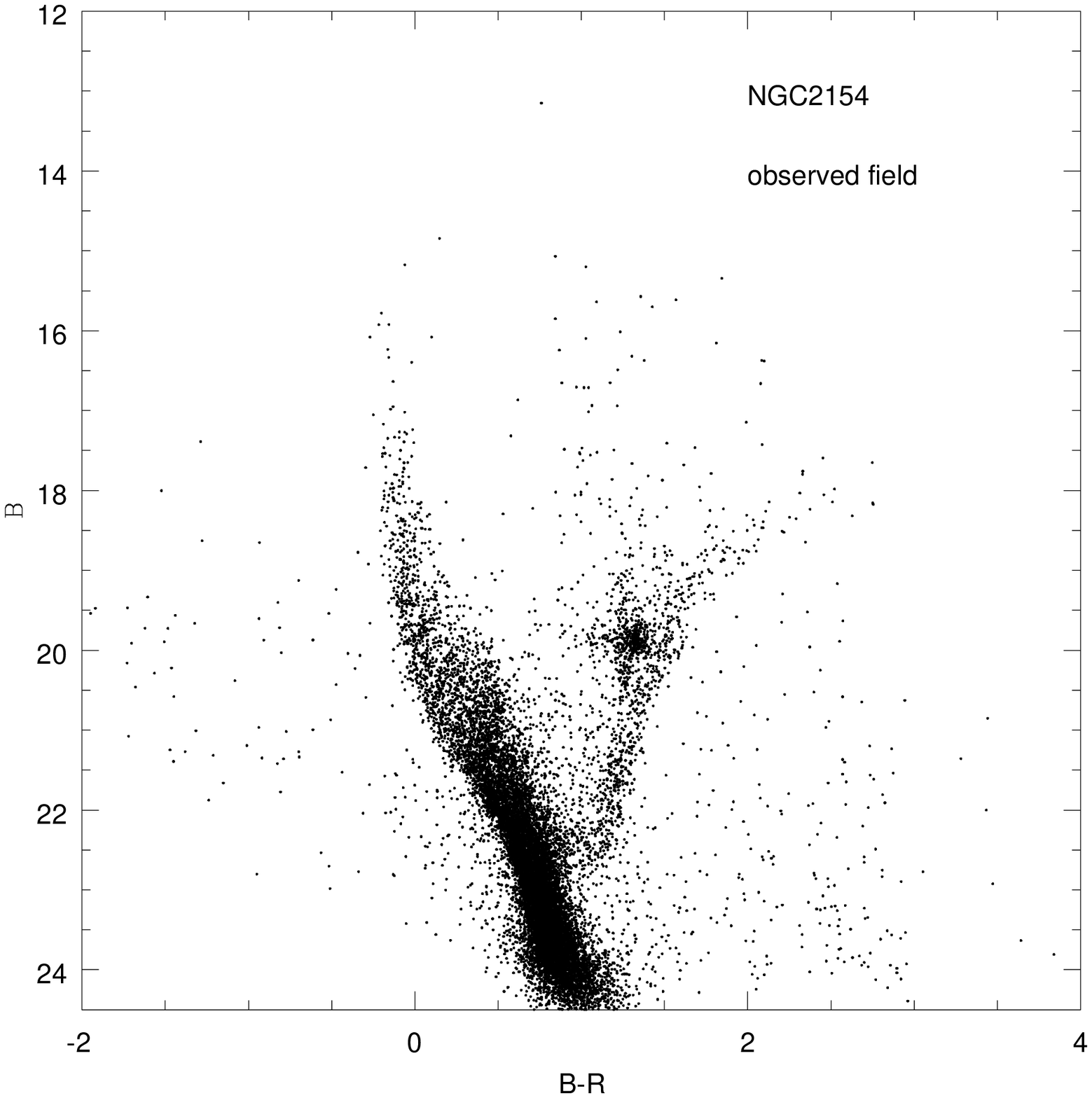}}\\}
\parbox{7.5cm}{\resizebox{7.5cm}{!}{\includegraphics{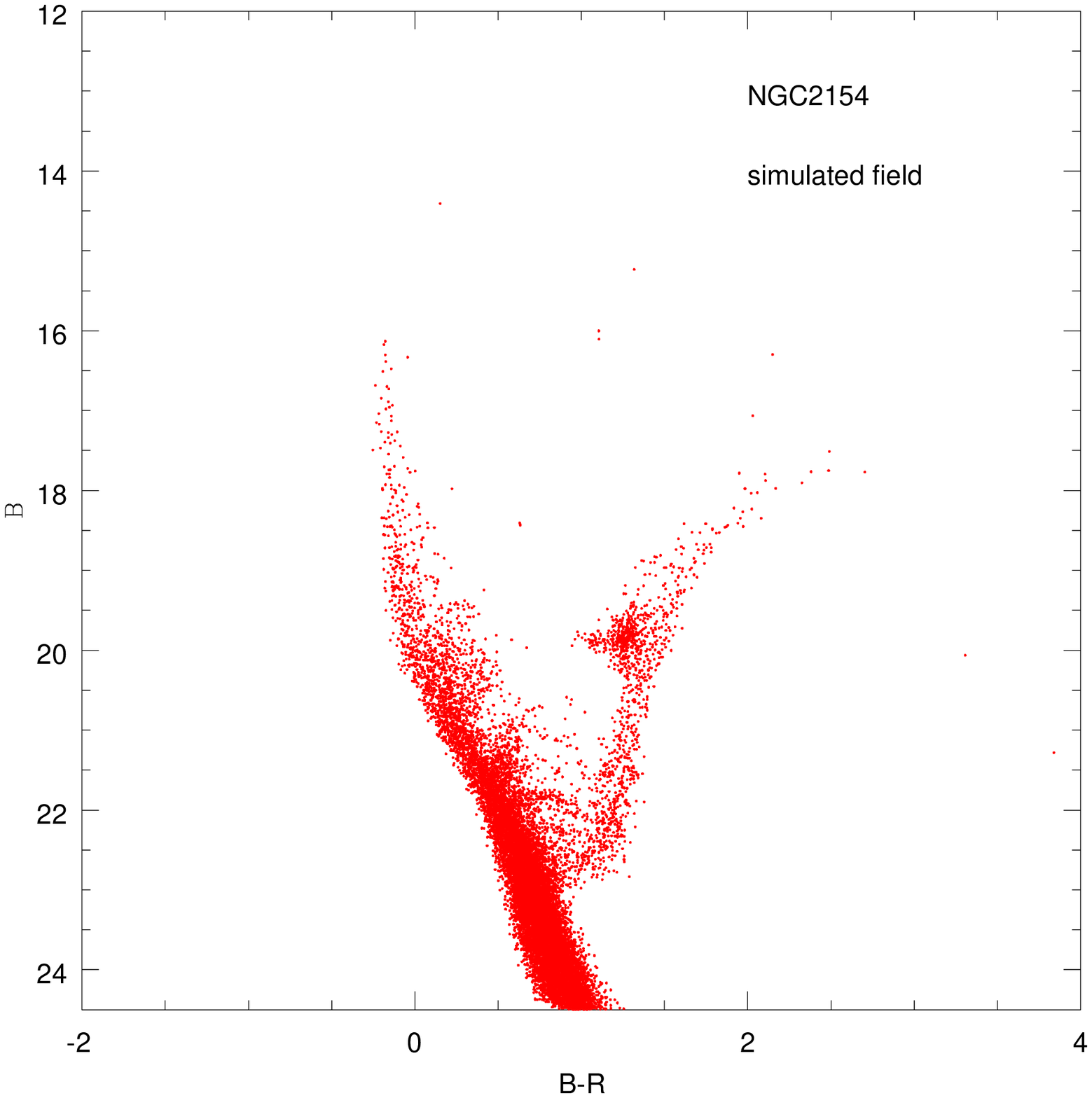}}\\}
\caption{Observed and synthetic CMDs of the field around NGC~2154. The
synthetic diagram is based on the SFR presented in Fig.~\ref{fig:ngc2154sfr}}
\label{fig:ngc2154cmd}
\end{figure*}

In Fig.~\ref{fig:ngc2154cmd} we compare the observed and synthetic CMDs of NGC~2154.\\

\subsection{NGC~1898}

NGC~1898 (= BSD99 2439) is located in the SW edge of the bar of the LMC ($\alpha=05^h 16^m 41.24^s$; 
$\delta=-69^{\circ} 39' 24.40''$). Because this field is more crowded than that of NGC~2154, and
because the observations were carried out in inferior seeing conditions, our data for NGC~1898 is
50\% complete at a brighter magnitude than that for NGC~2154 (50\% complete at $B = 20$ for the
cluster area, and at $B = 22.5$ for the field).\\

\begin{figure}
\centering
\resizebox{\hsize}{!}{\includegraphics{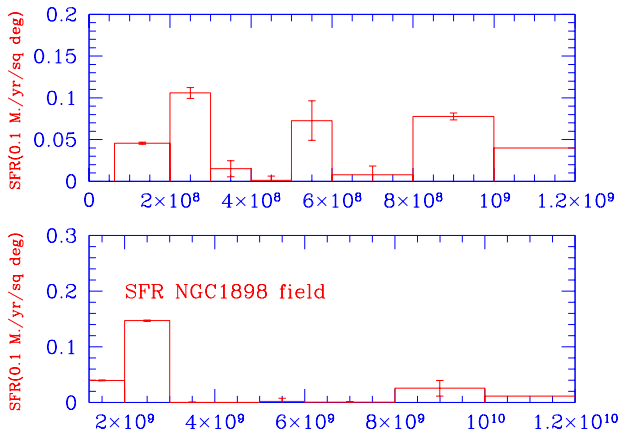}}
\caption{SFR in the field around NGC~1898.}
\label{fig:ngc1898sfr}
\end{figure}

As shown by Fig.~\ref{fig:ngc1898sfr}, our results indicate that the SFH in the field of NGC~1898 
had enhancements in the SFR at 200, 400, 800 Myr, similar to NGC~2154, with a notorious gap 4-5 Gyr. Although
not as notorious as in the case of NGC~2154, a shallow peak is also present at 6 Gyr (of the order of $0.002 M_{\odot}/yr/sqdeg$) and  consistent peaks at 2 and  8 Gyr. We stress that we are not precise at ages older than 6 Gyr.
  This
result is not inconsistent with those of \citet{olsen1999} for the LMC bar. Again, this peak and
the other bursts of star formation seen in Fig.~\ref{fig:ngc1898sfr} are probably the result of
dynamical interactions between the MCs, and of the MCs with the MW.\\
The peaks in the case of NGC~1898 are more consistent than in the case of NGC~2154 showing that the populations in the bar are much richer than those in the disk and obviously star formation is proportional to mass density. But the relative intensity of young ($<1Gyr$) and old ($>1Gyr$) populations shows that old populations in NGC~1898 are much more abundant than in NGC~2154 with respect to young populations. In fact for young population we have a factor 2 between NGC~1898 and NGC~2154 while for old population we have a factor 10. NGC~1898 is located in the bar while NGC~2154 is located in the disk. This is in agreement with Olsen (1999) that found a more conspicuous old component in the bar with respect to the disk.

\begin{figure*}
\centering
\hspace{0.2cm}
\parbox{7.5cm}{
\resizebox{7.5cm}{!}{\includegraphics{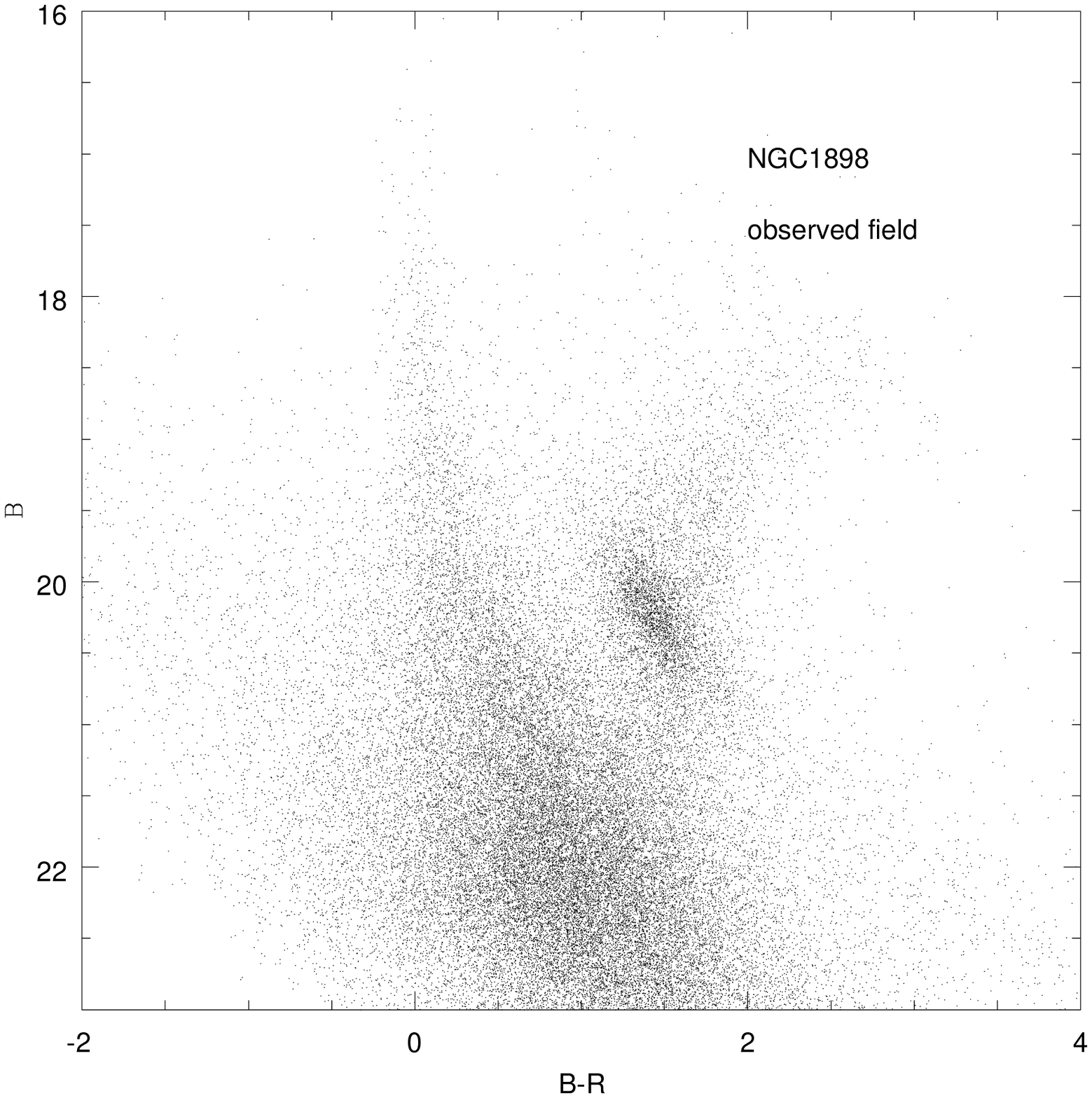}}\\}
\parbox{7.5cm}{
\resizebox{7.5cm}{!}{\includegraphics{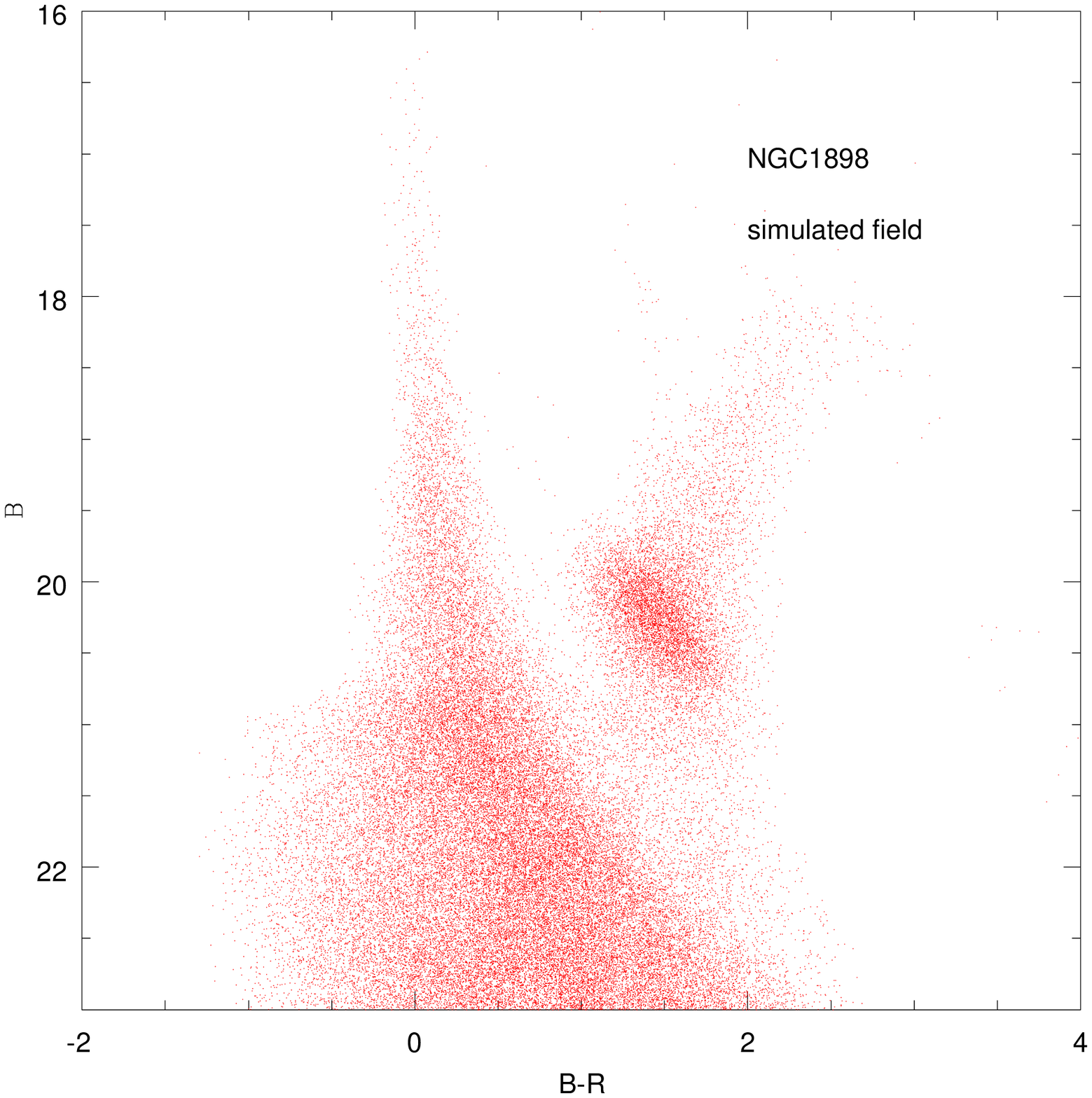}}\\}
\caption{Observed and synthetic CMDs of the field around NGC 1898. The
synthetic diagram is based on the SFR presented in Fig. \ref{fig:ngc1898sfr}}
\label{fig:ngc1898cmd}
\end{figure*}

In Fig.~\ref{fig:ngc1898cmd} we compare the observed and synthetic CMDs of NGC~1898.\\

\section{CLUSTER ANALYSIS IN THE FIELD OF NGC1898}\label{clu}
\begin{figure*}
\centering
\includegraphics[width=.45\textwidth]{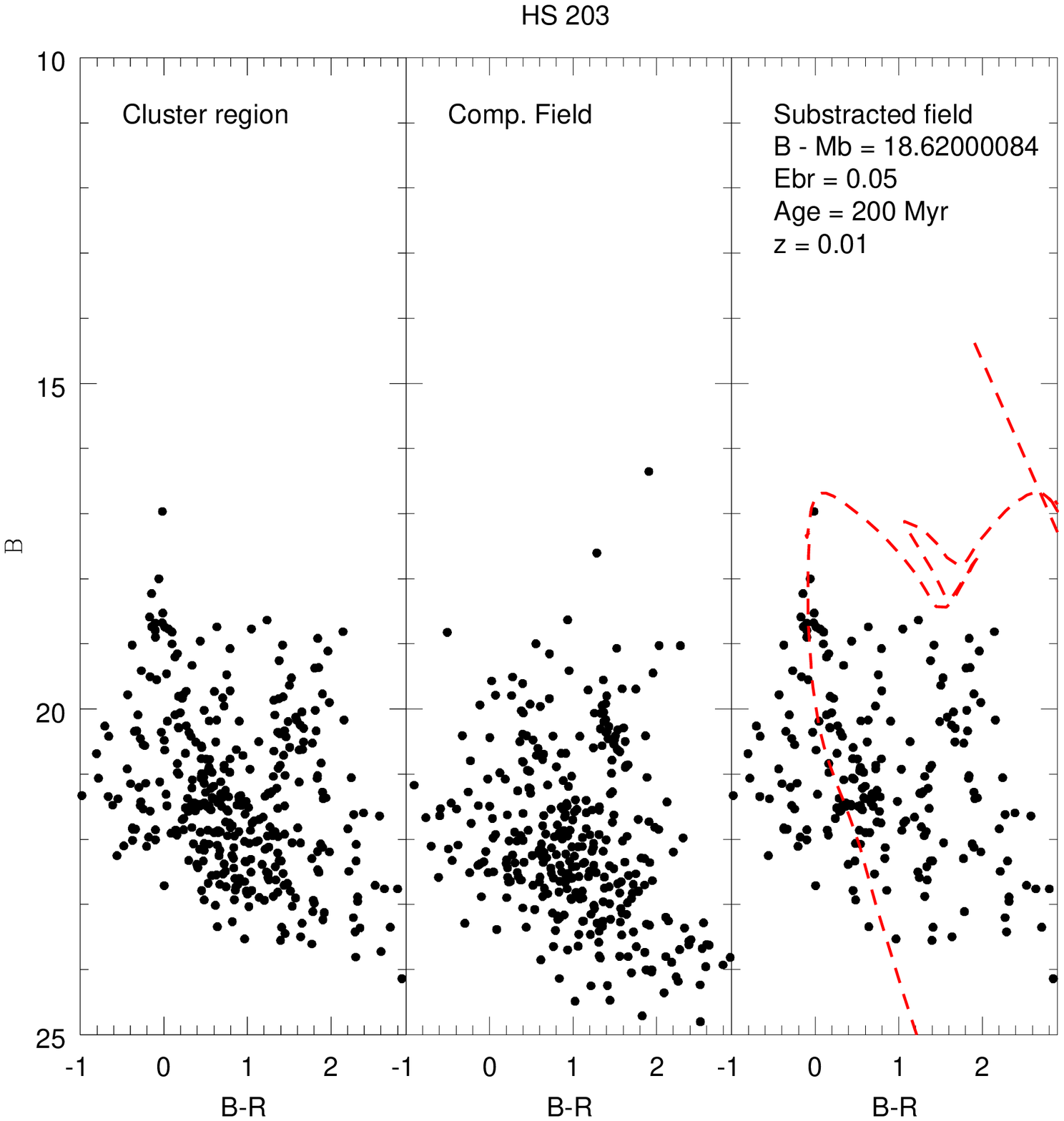}\includegraphics[width=.45\textwidth]{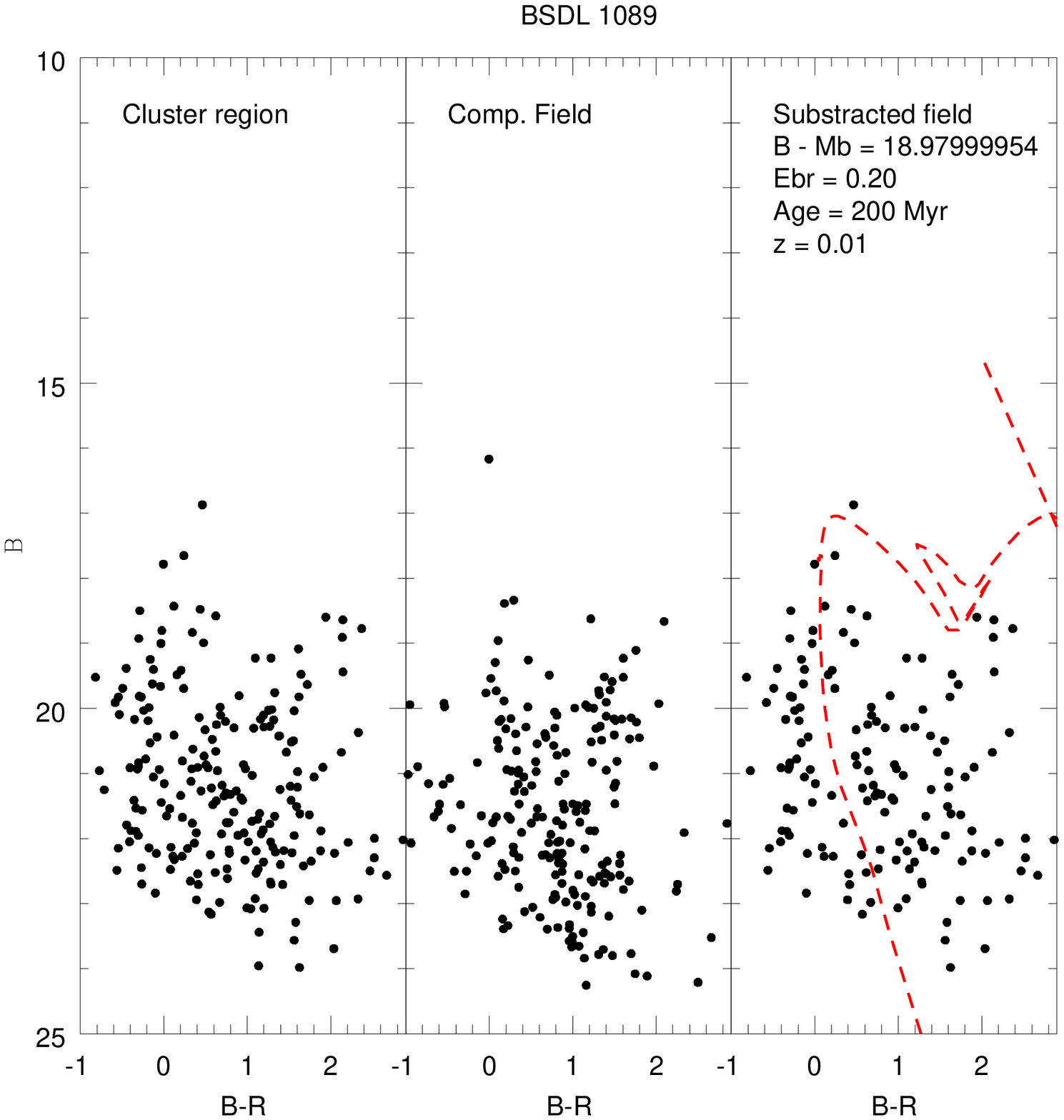}
\includegraphics[width=.45\textwidth]{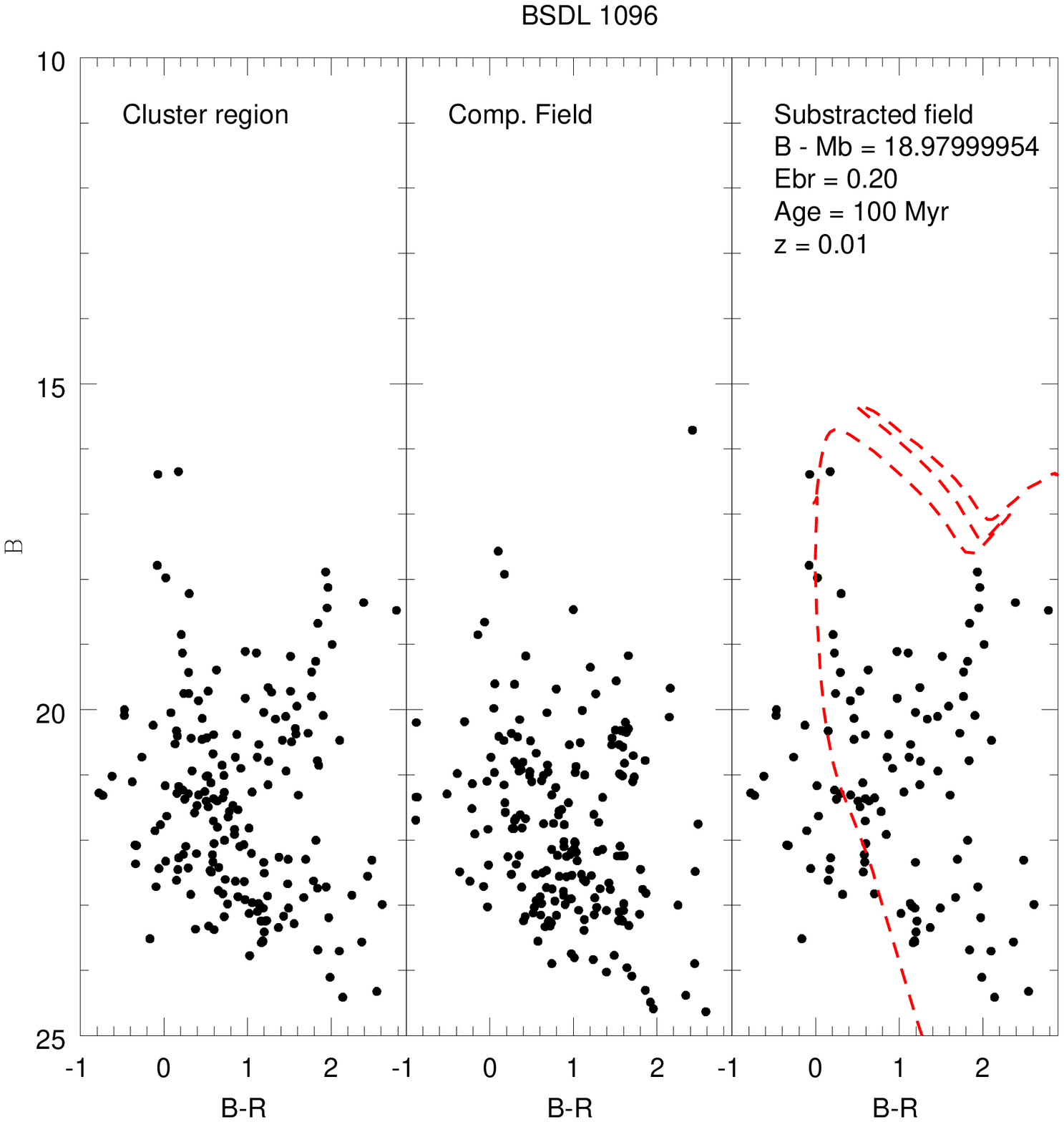}\includegraphics[width=.45\textwidth]{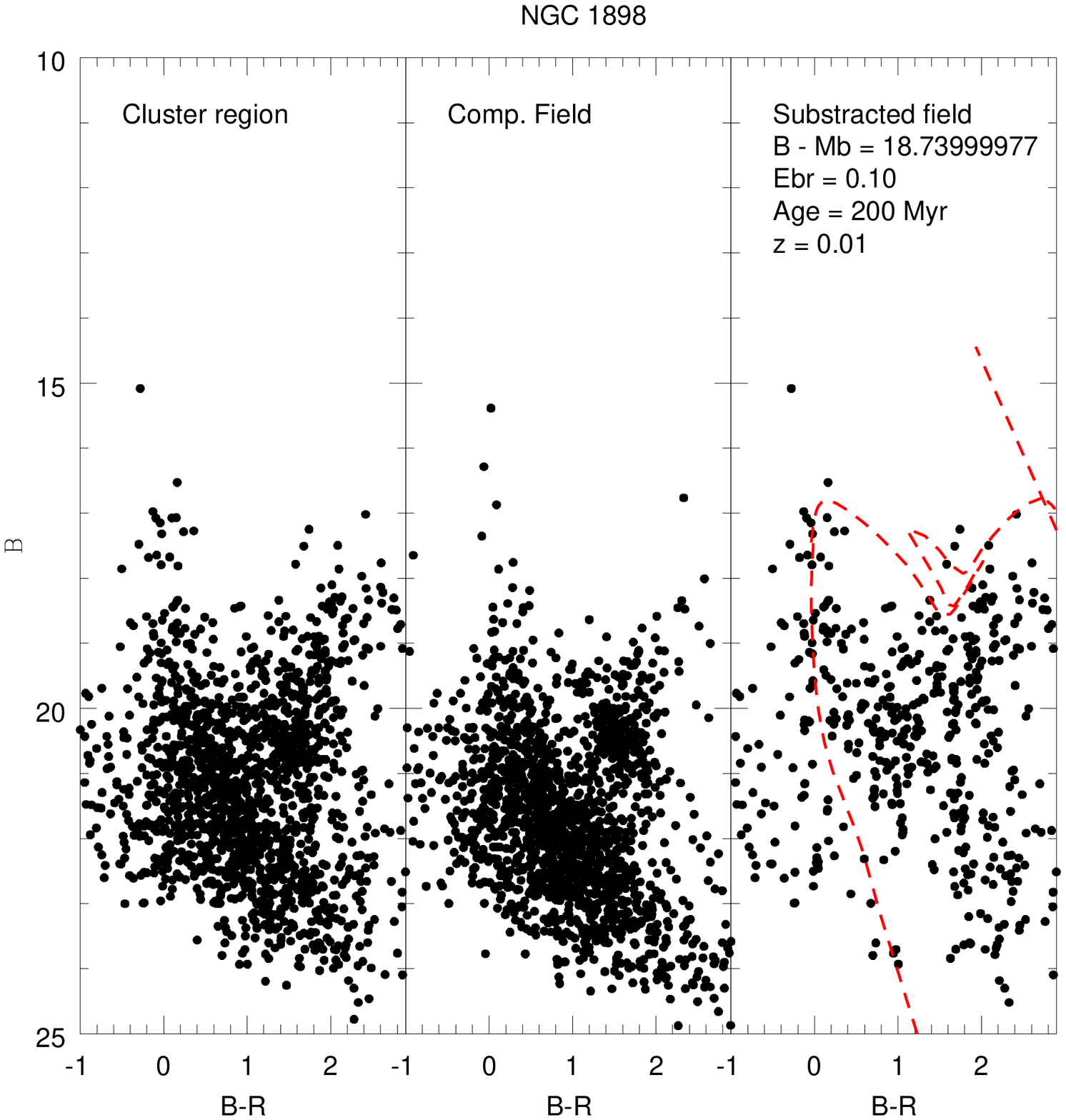}
\caption{Field star decontamination procedure. Left-hand panels are the CMDs of stars located in the
$cluster~regions$, central panels are the CMDs of stars in the corresponding $comparison~regions$, and
right-hand panels present the resulting $clean$ CMDs. Dashed lines are the best-fitting isochrones, from
Marigo et al. (2008). Cluster main parameters are indicated. For the diagrams of NGC~2154 refer to Bau07.}

\label{fig:cmds1}
\end{figure*}

\begin{figure*}
\includegraphics[width=.45\textwidth]{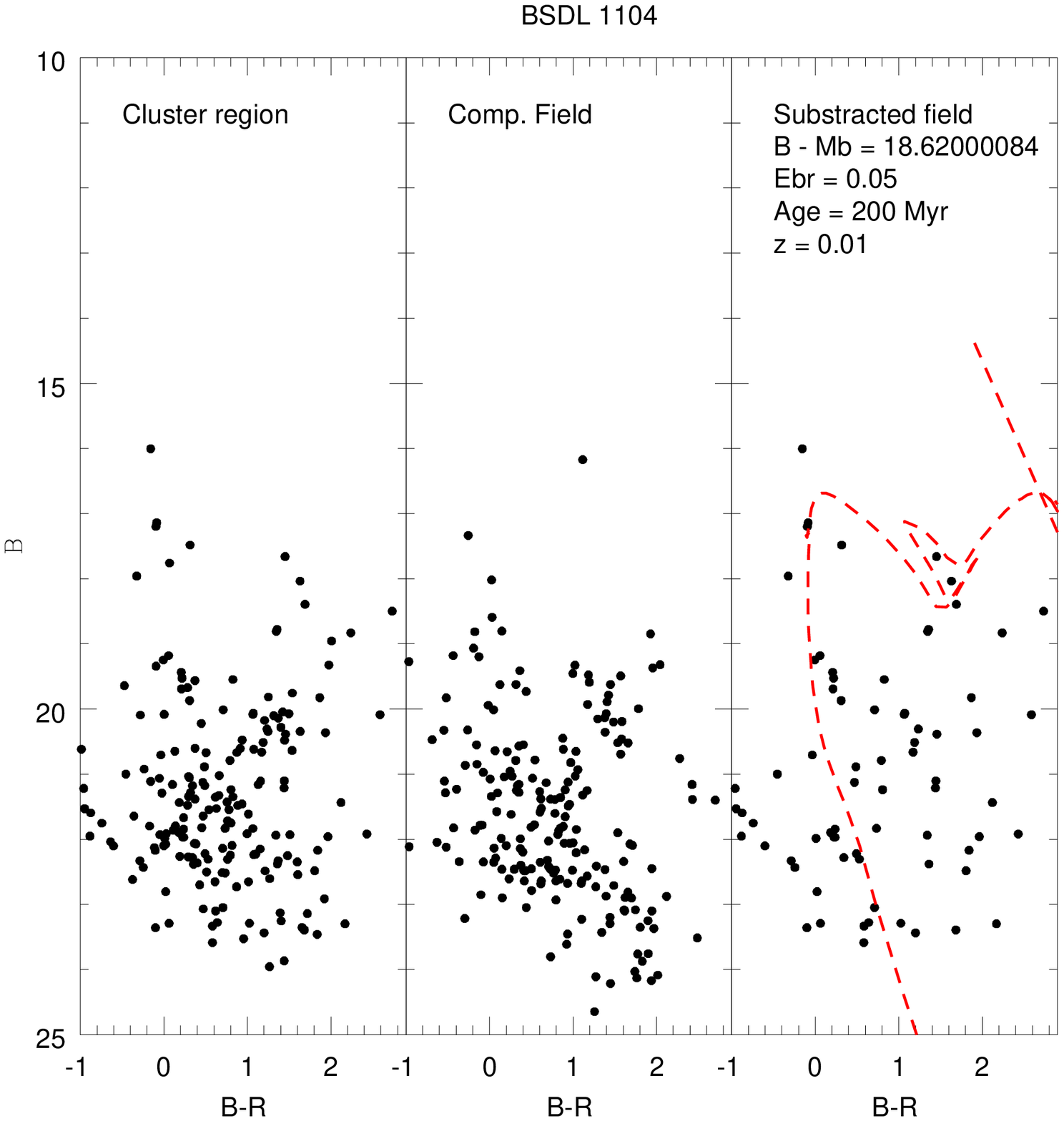}\includegraphics[width=.45\textwidth]{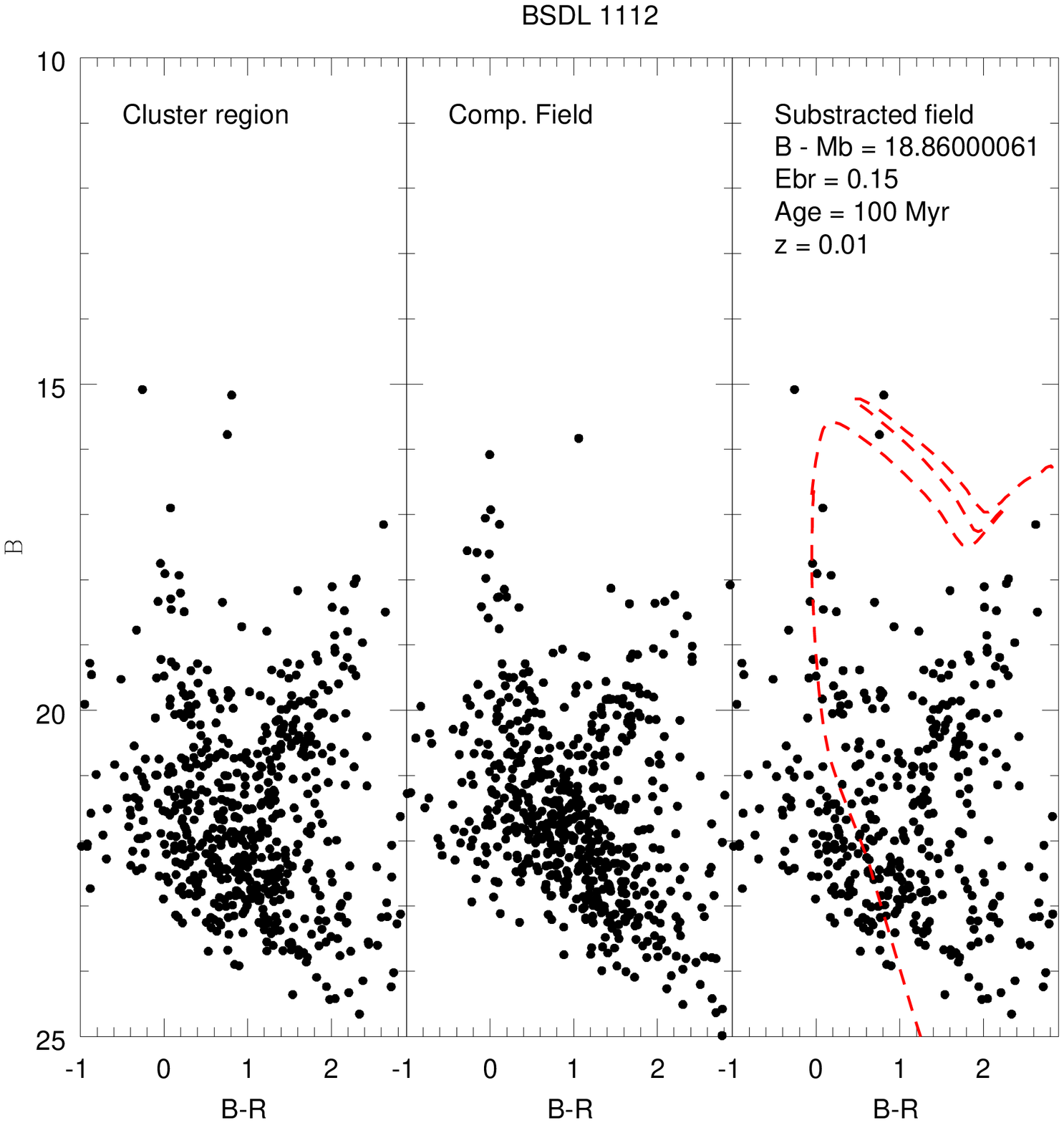}
\includegraphics[width=.45\textwidth]{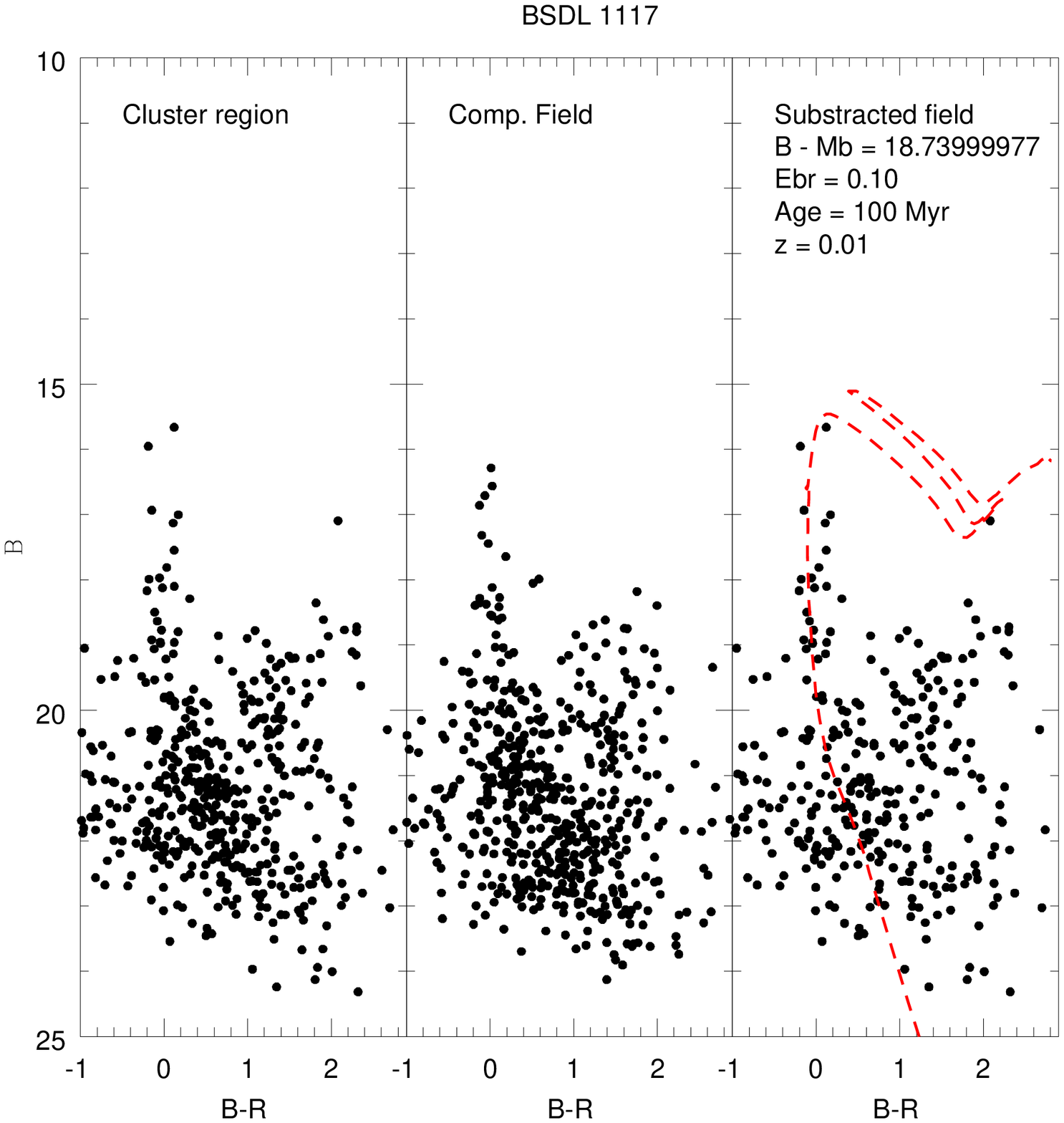}\includegraphics[width=.45\textwidth]{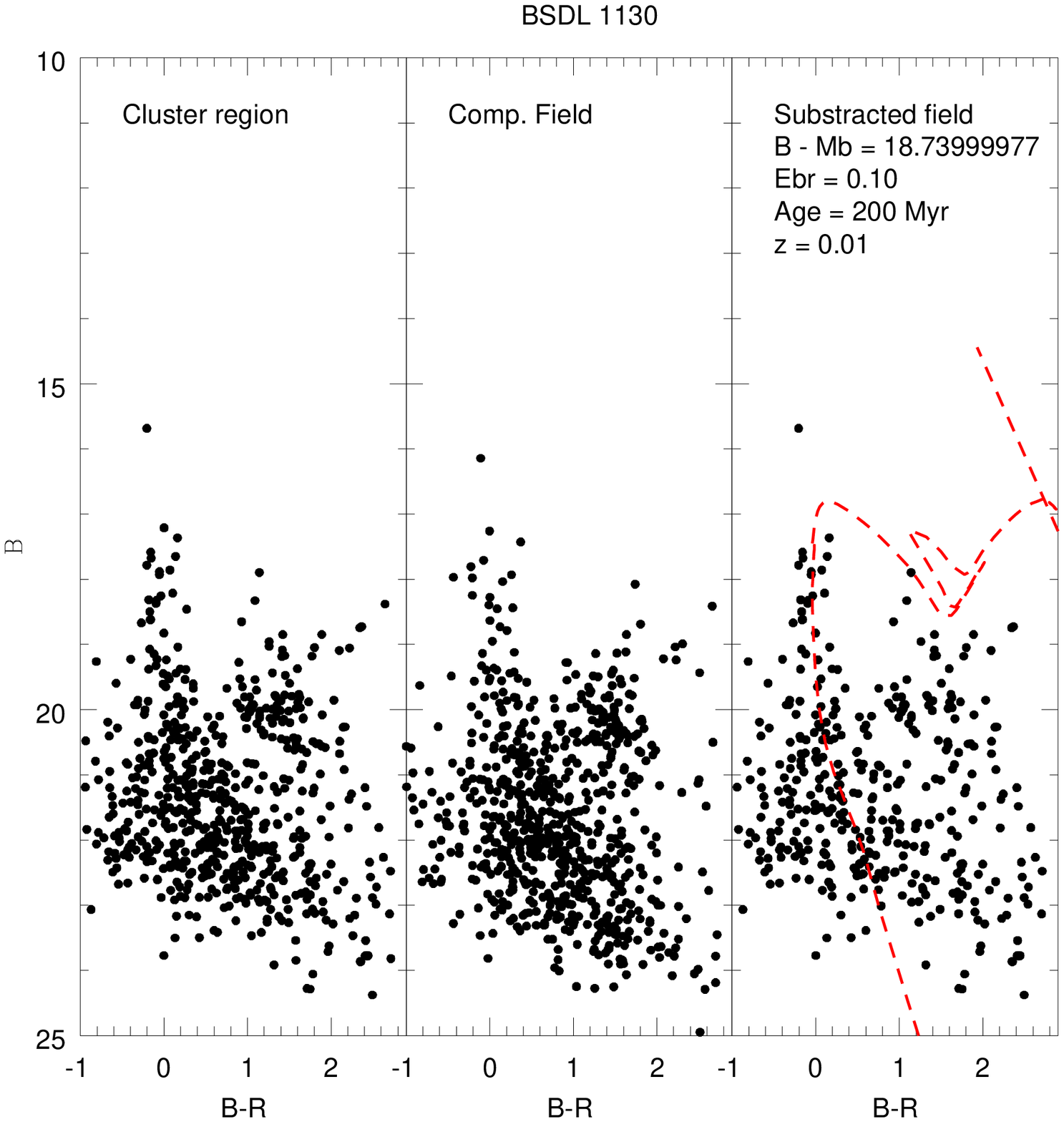}
\caption{Field star decontamination procedure. Left-hand panels are the CMDs of stars located in the
$cluster~regions$, central panels are the CMDs of stars in the corresponding $comparison~regions$, and
right-hand panels present the resulting $clean$ CMDs. Dashed lines are the best-fitting isochrones, from
Marigo et al. (2008). Cluster main parameters are indicated.For the diagrams of NGC~2154 refer to Bau07.}
\label{fig:cmds}
\end{figure*}

The analysis of NGC~1898 field showed the presence of small clusters. In the followings first we describe the procedure of determination of cluster parameters then we discuss the relation with the field.\\
 \subsection{CLUSTER PARAMETERS}
The determination of cluster centers and radii was different for NGC2154 and for the other clusters in NGC1898 area. In the first case the determination of the center was done by visual inspection as the position of the peak density while the determination of the radius was done through the fit with Elson profiles (see Bau07).
  For the other small clusters in the NGC1898 area we tested several centers until
we obtained the most clear case for a stellar concentration at the center, and adopted as cluster
center the position of the peak density. The adopted sizes correspond to radii where the densities 
get confused with the background level. In all this procedure, we considered only the brightest 
stars ($B < 21$) in order to avoid noise contamination by spurious detections. The resulting
center coordinates and radii are depicted with circles in Fig. 1 and given numerically in 
Table~2. The data presented for NGC~2154 in this table are from
Bau07.

\begin{table*}
\label{tab:param}
\begin{center}
\begin{tabular}{ccccccl}
\hline
Name  & $\alpha_{2000}$ & $\delta_{2000}$ &$Id_{B99}$ & $R[']$ & $Age[Myr]$& $z$ \\
\hline
 HS~203    & 05:16:14.0 & -69:49:32.0 &2408 & 0.8  $\pm 0.1$ &  200 $\pm 40$  & 0.010 \\
 BSDL~1089 & 05:16:21.0 & -69:36:03.0 &2418 & 0.6  $\pm 0.2$ &  200 $\pm 40$  & 0.010 \\
 BSDL~1096 & 05:16:26.0 & -69:40:28.0 &2428 & 0.55 $\pm 0.1$ &  100 $\pm 20$  & 0.010 \\
 NGC~1898  & 05:16:42.0 & -69:39:22.0 &2439 & 1.6  $\pm 0.2$ &  200 $\pm 40$  & 0.010 \\
 BSDL~1104 & 05:16:43.0 & -69:35:47.0 &2443 & 0.55 $\pm 0.1$ &  200 $\pm 40$  & 0.010 \\
 BSDL~1112 & 05:16:57.0 & -69:40:31.0 &2464 & 1.0  $\pm 0.1$ &  100 $\pm 20$  & 0.010 \\
 BSDL~1117 & 05:17:10.0 & -69:35:34.0 &2473 & 0.9  $\pm 0.1$ &  100 $\pm 20$  & 0.010 \\
 BSDL~1130 & 05:17:28.0 & -69:35:38.0 &2498 & 1.0  $\pm 0.2$ &  100 $\pm 20$  & 0.010 \\
 NGC~2154  & 05:57:38.2 & -67:15:40.7 &6351 & 1.73 $\pm 0.2$ & 1700* & 0.005 \\
\hline
\end{tabular}
\end{center}
\caption{Parameters of the clusters investigated.We give the name, the position, the radius in arcmin, the age and the metallicity. 
$Id_{B99}$ is simply the running number
in the catalogue of Bica et al. 1999.* cluster NGC~2154 presents a superposition of three stellar populations as discussed in Bau07.} 
\end{table*}

We then constructed $clean$ color-magnitude diagrams (CMDs) of the clusters by selecting all stars 
located in each $cluster~region$, and in a $comparison~region$ of identical size, and statistically 
subtracting the latter from the former. This field star decontamination method is described in
detail by \citet{vallenari1992} and by \citet{gallart2003}.
In Fig.~\ref{fig:cmds} we present the result of carrying out this procedure on NGC~1898 and the
seven small clusters. An equivalent result for NGC~2154 can be found in Bau07 (see their
Figs. 8 and 9).\\

We are going to consider all the objects as genuine star clusters,
since they are made of a significant stars overdensity, and the stars
producing the overdensity exhibits in the CM distintive features.\\
The bright part of each $clean$ CMDs ($B < 19$) was then compared with theoretical stellar evolutionary 
models from the Padova group (\citet{marigo2008}) (see Fig.~\ref{fig:cmds1} and \ref{fig:cmds}), thus obtaining the age
estimations indicated
in Table~\ref{tab:param}. For all clusters in the region of NGC~1898 (see Fig.~\ref{fig:dss}), 
we adopted a metallicity for the LMC of $[Fe/H] = -0.30$, in agreement with \citet{rolleston2002},
which corresponds to $z = 0.010$. The $B-M_B$ and $E_{B-R}$ values for each cluster are indicated in
Fig.~\ref{fig:cmds1} and \ref{fig:cmds}. The are consistent with a distance modulus of $B_0-M_B = 18.5$, in agreement with
\citet{westerlund1997}. For the region of NGC~2154 we adopted the
parameters indicated in Bau07. Some of the clusters like BSDL~1104 and BSDL~1096 are uncertain as they show few stars
in the subtracted CM diagram.\\

Two distinct populations can be noticed: an older one (1-2 Gyr) exemplified by NGC~2154, and a younger
one (100-200 Myr), exemplified by NGC~1898 and the seven small clusters.  As explained in Section 4,
we understand these two distinct populations as the result of bursts of star formation in the LMC.\\
\subsection{CLUSTER POPULATION} 
The cluster analysis showed the presence of a coeval young (\~100-200Myr)
 population of clusters spread around NGC~1898. This population corresponds to a peak in the SFH of NGC~1898 field. 
The percentage of star formation happening in cluster relative to that taking place in the field is  investigated 
by comparing the peak in the distribution of the global field to that of the subtracted field. The result does not show 
any significant change in the peak value
 possibly meaning that the SF at young ages takes place all over the area and is not concentrated on the cluster sites.\\
We notice how the cluster and field populations are coeval. The field subtraction is a critical point in the definition of
 cluster ages through isochrone fitting. We recall that we cannot age older components in the clusters due to the depth of 
the photometry. So we probably identify the youngest episode of star formation happening both in clusters and field.\\

\section{CONCLUSIONS}\label{conclusions}

In this paper we investigated nine clusters of the LMC (NGC~2154, NGC~1898, and seven small 
clusters in the vicinity of the latter), and their related fields.\\

Two distinct populations of clusters were found: one cluster (NGC~2154) has a mean age of 1.7 Gyr,
with indication of extended star formation over roughly a 1 Gyr period (Bau07), while all others
have ages between 100 and 200 My.

We also derived the SFRs for their adjacent fields. In the case of the NGC~2154 field, enhancements
in the SFR are seen at 200, 400, 800 Myr; and in the case of the  NGC~1898 field at 1, 6, 8 Gyr,
with a notorious gap 4-5 Gyr. This implies that SFH proceeded in
somewhat different ways in the two regions.

These bursts of star formation seem to be consistent with the dynamical interactions believed to
have occurred between the LMC and the SMC at 200 Myr, and between the MCs and the MW at 1.5 Gyr.

\section{Acknowledgment}

EC and RAM acknowledge support by the Fondo Nacional de Investigaci\'on Cient\'{\i}fica y 
Tecnol\'ogica (proyectos Fondecyt No. 1050718 y No. 1110100), the Chilean Centro de
Astrof\'{\i}sica (FONDAP No. 15010003) and the Chilean Centro de Excelencia en Astrof\'{\i}sica y 
Tecnolog\'{\i}as Afines (PFB 06).
         
\bibliographystyle{mn2e}
\bibliography{biblio}
\end{document}